\documentstyle{article}
\title{
\bf 
2D Lorentzian Gravity as \\
2D Euclidean Gravity with Ising spins\\
}
\author{ W. Beirl\\
         Institut f\"ur Kernphysik\\
         Technische Universit\"at Wien\\
         A-1040 Vienna\\
         Austria\\
         \\
         {\it and}\\
         \\
         D.A. Johnston \\
         Dept. of Mathematics\\
         Heriot-Watt University\\
         Riccarton\\
         Edinburgh, EH14 4AS\\
         Scotland}
\begin{document}
  \maketitle
                      {\Large
                      \begin{abstract}
%
We suggest a generalization of the dynamical triangulation
approach to quantum gravity with both timelike and spacelike edges,
which can serve as a toy model for quantum gravity in the Lorentz
sector in two dimensions. It is possible to consider
the model in a purely Lorentzian sector or to
relax this constraint and allow local signature changing moves.
We show that, with suitable
conventions, the model is equivalent to
an Ising model coupled to 2D Euclidean quantum gravity
and conduct a preliminary numerical simulation
of the Lorentz sector.
\\
\\
\\
\\
%
                        \end{abstract} }
%
  \thispagestyle{empty}
%
%
  \newpage
%
                  \pagenumbering{arabic}

\section{The Model}
Two schools currently exist for the
numerical exploration of quantum gravity
which are both based on simplicial discretization of spacetime.
In the dynamical triangulation approach
\cite{0} the edge lengths on the
discretized lattice are fixed and connectivities are allowed to vary whereas
in the Regge calculus approach \cite{1},
at least insofar as practical implementations are concerned,
the connectivities are fixed and edge lengths are allowed to vary.
In two dimensions dynamical triangulations are a
direct implementation of matrix models and the results
obtained are in good agreement with the theory,
whereas there is currently some uncertainty
as to whether the Regge calculus agrees with
analytical results \cite{2}. 
Indeed, some recent results raise the question
of whether two dimensions is
a suitable testing ground for discretized
quantum gravity at all \cite{BB}. In higher dimensions, however,
the Regge calculus and dynamical triangulations appear to be
in at least qualitative agreement on the phase structure
of quantum gravity \cite{2a,2b}.

All of the work alluded to above has explored discretized gravity
in the Euclidean sector and its relation to Lorentzian gravity
is a moot point, as is also the case in the continuum theory.
In principle the Regge approach is not shackled to Euclidean space.
The volume of a Regge simplex in $n$ dimensions is given by
\begin{equation}
V_n = { 1 \over n!} \sqrt{ \det (\vec e_i \cdot \vec e_j )}
\end{equation}
where the vectors $\vec e_i$ start at some basis vertex, say $0$,
and point to vertex $i$ thus spanning the simplex. In the two dimensional 
case the volume formula gives the
the area of a triangle as
\begin{equation}
A = { 1 \over 2} \left| \begin{array}{cc} q_1 &{1 \over 2} (q_1 + q_2 - q_0) 
\\
                               {1 \over 2} (q_1 + q_2 - q_0) & q_2
             \end{array} \right|^{1/2}
\label{area}
\end{equation}
where $q_0,q_1,q_2$ are the squared edge lengths of the triangle.
The triangle can be Lorentzian if the determinant
is less than zero or
Euclidean if it is greater than zero. In simulations to date
using Regge calculus the edge lengths have been generally
taken to be continuous variables with triangle inequalities
being explicitly enforced to ensure that the triangles stay Euclidean.
One interesting alternative scheme that has been proposed is the
so-called Ising link quantum gravity in which only two
squared edge lengths are allowed, $q = 1 + \epsilon \sigma$,
where $\sigma$ is an Ising variable and $\epsilon$ is
chosen to maintain Euclidean triangles ($\epsilon<3/5$ in two
dimensions) \cite{3}.

In the spirit of the Ising link approach, we consider a
generalization of dynamical triangulation with
both spacelike and timelike edges, $q_{space}=+1, q_{time}=-1$.
We can accommodate both Euclidean and Lorentzian triangles
in this scheme: triangles with
$( +1, +1, +1 )$ and $( -1, -1, -1 )$ edges give Euclidean areas
from equ.(\ref{area}); whereas triangles with either $( +1, -1 , -1 )$
or $( +1, +1, -1 )$ edges give Lorentzian areas.
It is possible to take the edge
lengths squared themselves as Ising spins, but
instead we will take them to be 
the product of Ising spins at the end of each link, which will allow us to derive some
properties of 
our toy model of quantum gravity in the Lorentz sector from already
solved Ising models in the Euclidean sector.

If we allow all four possible sorts
of triangle and assign fugacities $g$ to the Euclidean triangles
and $z$ to the Lorentzian triangles we can write down
the following partition function for our model
\begin{equation}
Z = \sum_{\Delta} \sum_{configs} g^{(\sharp \; Euclidean \; \Delta s)} \;  z^{(  \sharp \;
Lorentzian \; \Delta s)}
\end{equation}
where the outer sum runs over (Euclidean) triangulations and the inner sum is 
over
the different possible ways of soldering the various sorts of triangles 
together.
This partition function can be retrieved as the free energy
of a two-matrix model \cite{4a} whose perturbative expansion
generates our triangulated universe of mixed
Lorentzian and Euclidean patches
\begin{equation}
U =   tr \left( {1 \over 2 } S^2 + { 1 \over 2 } D^2
 -{ g\over 3}   S^3  - { g \over 3} D^3 -  z S D^2  -z D S^2 \right)
\label{action0}
\end{equation}
where we use the $M \times M$ hermitian matrix
$S$ to represent a spacelike edge
and the $M \times M$ hermitian matrix
$D$ to represent a timelike edge.
The $S^3$, $D^3$ terms thus represents the Euclidean triangles,
the $S D^2$, $S^2 D$ terms the Lorentzian triangles and the $S^2$ and $D^2$
terms solder together different triangles.

As it stands this model has
not been solved
\footnote{A possible method for finding a solution
in such models
has been sketched recently in \cite{4a}. It is also
worth noting that making a field redefinition 
for the Ising model with an external field on $\phi^3$ graphs
gives an action with identical terms to equ.(\ref{action0})
but the coefficients of $S^3$ and $D^3$ are not equal, 
nor are those in front of $D S^2$ and $S D^2$.}
so we make a further bold simplification
and restrict ourselves
by fiat
to one sort of Euclidean triangle, namely $( +1, +1, +1 )$,
and one sort of Lorentzian triangle, $( +1, -1 , -1 )$
which leaves us with the following matrix model
\begin{equation}
U =   tr \left( {1 \over 2 } S^2 + { 1 \over 2 } D^2
 - { g \over 3 } S^3  -  z S D^2 \right).
\label{action}
\end{equation}
This two matrix model is known to represent
the Ising model on dynamical triangulations \cite{4a,4}
with the spins at the triangle vertices,
rather than the more usual formulation
with the spins at triangle centres \cite{5}.
In this context,
$S$ represents edges with similar spins at each end
and $D$ represents edges with differing spins at each end.
In Figures.1-4 we denote the $S$ edges by bold lines and the $D$
edges by dotted lines, as well as indicating the corresponding
vertex spin arrangements.
After suitable scalings the Ising action on dynamical triangulations
can be written as
\begin{equation}
U =   tr \left( {1 \over 2 } S^2 + { 1 \over 2 } D^2
 - { g \over 3 } S^3  -  g \exp ( - 2 \beta ) S D^2 \right).
\label{actionI}
\end{equation}
so we see that the zero temperature ($\beta \rightarrow \infty$)
limit is the Euclidean sector when the model
is regarded as Ising link quantum gravity, whereas an
antiferromagnetic limit ($\beta \rightarrow - \infty$, $g
\rightarrow 0$ with $z=g \exp ( - 2 \beta )$ finite)
can be regarded as the purely Lorentzian limit.
In this limit the action becomes
\begin{equation}
U =   tr \left( {1 \over 2 } S^2 + { 1 \over 2 } D^2 - z S D^2 \right).
\label{actionII}
\end{equation}
and the gaussian integration over $S$ may be performed, giving
\begin{equation}
U =   tr \left( {1 \over 2 } D^2 - { z^2 \over 2} D^4 \right). 
\label{actionIII} 
\end{equation}
which is again a pure gravity action. In geometrical terms this
means that the triangulations generated
in the Lorentzian limit that gives equ.(\ref{actionII}) 
are those which admit a perfect matching in which the
triangles are paired off into squares \cite{bachas}. This  
is clear from equ.(\ref{actionIII}) where
the effect of integrating out the $S$ is to
give a $D^4$ term that generates squares.

If we take the opposite convention for 
the Lorentzian triangles and keep 
$( +1, +1, +1 )$ and $( +1, +1 , -1 )$
triangles only we have the action
\begin{equation}
U =   tr \left( {1 \over 2 } S^2 + { 1 \over 2 } D^2 
 - { g \over 3 } S^3 - z D S^2 \right).
\label{actionIV}
\end{equation}
which is is now gaussian in $D$, allowing it to be integrated out
to give
\begin{equation}
U =   tr \left( {1 \over 2 } S^2  - { g \over 3 } S^3 - {z^2 \over 2}
S^4 \right).
\label{actionV}
\end{equation}
This is an
action for a universe built out of both triangles
and squares. This action is known to give the Yang-Lee edge
singularity on a random surface for suitably tuned $g$ and $z$
\cite{staud}, but
generically it will be pure gravity. Again the
purely Lorenztian limit is to take $g \rightarrow 0$
which leaves us with a theory of squares alone (but with 
spacelike edges, rather than the timelike edges of
equ.(\ref{actionIII})).

We have glossed over one important respect in which our toy model fails to be
realistic: we have implicitly assumed
that $g$ and $z$ are {\it real} to get the standard critical behaviour,
ie $g = \exp ( - \mu)$, $z =  g (- 2 \beta) $, so that
for a fixed topology
\begin{equation}
Z \simeq \sum_n  \exp ( - \mu n ) \cdots
\end{equation}
where $n$ is the number of vertices in a triangulation. In the continuum
limit the exponential gives the cosmological constant
term $\int \sqrt{g}$ without a factor of $i$ in front, whether
we are in the Euclidean sector or the dense Lorentzian sector 
\footnote{The 
curvature term, $N^{\chi}$, that has been suppressed at fixed topology 
in the above also fails to display a factor of $i$
in the continuum limit, giving $\int \sqrt{g} R$.}.
However, it is known that the action in equ.(\ref{actionIII})
may be defined for complex $z$ \cite{david}, so we can approach
the standard continuum limit along a complex path, which is
precisely what is required for obtaining a factor of $i$ in the action.
The action for the Yang-Lee edge singularity in equ(\ref{actionV}) may also be
analytically continued to complex couplings.
In any event, these problems are swept under the carpet in a 
microcanonical simulation at fixed topology as the ``factors of $i$'' 
only come into play when the number of vertices or the topology
are changed. The problem re-emerges when 
one considers applying a scaling analysis
to simulations of differently sized lattices.
It is interesting to 
note in passing that a rather more sophisticated
form of Regge calculus with more than two quantized edge lengths devised
by Ponzano and Regge for three dimensional gravity \cite{RP} also fails
to reproduce the correct factors of $i$. 
When the continuum 
limit is taken in the Lorentzian sector
such a theory gives no $i$ in front of the action \cite{BF}, whereas
taking the continuum limit in the Euclidean sector does
give a factor of $i$.

\section{A Numerical Simulation}

In a numerical
simulation of the model
the allowed moves
for various sorts of edges can be deduced
by looking at the Ising interpretation of the model.
One new  
feature compared with an undecorated
Euclidean triangulation is that the standard ``flip'' move used in simulations
can have the effect of changing the sign of an edge, depending
on the local vertex spin configuration, which in turn has the effect of
changing the signature of one of the participating triangles.
The model thus incorporates local signature changing moves.
The choice of only $( +1, +1, +1 )$ triangles
is completely trivial in a purely
Euclidean theory - it is simply a matter of convention.
In a purely Lorentzian sector we can see
that the choice of $( +1, -1 , -1 )$ triangles
is compatible with the dynamical triangulation
moves shown in Figure.1, neither
of which change the sign of the flipped edge
or the nature of the participating triangles.
The further two flip moves
shown in Figure.2 {\it do} change the
sign of the flipped edge and exchange Lorentzian
and Euclidean triangles. These signature changing
moves will be present if we choose a non-zero
fugacity $g$ for the Euclidean triangles.

As we have defined the signature of an edge in terms
of the Ising spins living at either end it is also
possible for a changing spin to influence the local geometry.
An example of this in the purely Lorentzian sector
is shown in Figure.3 where flipping the central spin
still keeps only $S D^2$ triangles but
rotates the bold $S$ edges around the central point.
In general any vertex with an equal number
of alternating $S$ and $D$ edges emanating from it
allows a similar sort of move, whilst still keeping
the restriction to $S D^2$ triangles only.
In all cases, as we might expect given
the underlying matrix model of equ.(\ref{action}),
$( +1, +1, -1)$ and $(-1, -1, -1)$ triangles
are consistently excluded, if the starting configuration 
does not contain them. For a lattice of toroidal
topology, for instance, a section of a suitable
starting configuration might look like Figure.4
in which horizontal spacelike lines are joined by the
inclined timelike edges. 
Such a starting configuration gives a foliation
of the spacetime manifold by spacelike lines.
A lattice of spherical
topology would require in addition ``caps'' of 
purely timelike edges radiating from the poles.

We 
conducted a simulation 
of the model with the action in equ.(\ref{actionII})
(``perfect matching'') 
which is in a purely Lorentzian sector
\footnote{In this sector there are no signature changing moves, 
only those in Fig.1 being admissable.} 
according to our definitions and
measured the probability $\omega(n)$ 
for the occurence of vertices with coordination number $n$. 
These are
listed in the following table for $n=3, 4, .. 10$ for a 
toroidal lattice with 1024
vertices and a statistics of 40k {\it independent} configurations
obtained via 1M sweeps. The statistical error is about $0.2$ percent. 
The columns $\omega_L$ and $\omega_E$ contain the numerical values for Lorentz
and Euclidean sector respectively, 
while $\omega_{exact}$ shows the exact result for the Euclidean sector. 

\begin{center}
\begin{tabular}{|l|l|l|r|}\hline
$n$  &  $\omega_L$ & $\omega_E$ & $\omega_{exact}$ \\[.05in] 
\hline 
$3$    & 0.193 & 0.210 & 0.210937... \\
$4$    & 0.212 & 0.197 & 0.197753... \\
$5$    & 0.156 & 0.155 & 0.155731... \\
$6$    & 0.119 & 0.116 & 0.116798... \\ 
$7$    & 0.086 & 0.086 & 0.086034... \\ 
$8$    & 0.063 & 0.062 & 0.062912... \\ 
$9$    & 0.045 & 0.045 & 0.045873... \\ 
$10$   & 0.033 & 0.033 & 0.033422... \\[.05in] \hline

\end{tabular}
\end{center}
\vspace{.1in}
\centerline{{\bf Table 1:} Measured
values of the probability $\omega(n)$
for the occurence}
\centerline{of vertices with coordination number $n$.}
\vspace{.1in}
\noindent
We can see that the main effect in the Lorentz sector
is to increase $\omega(4)$ 
(and to a much lesser extent $\omega(6)$) at the expense of $\omega(3)$,
the $\omega$'s for larger $n$ being essentially identical.
This ties in with the intuition that a perfect matching
is going to be easier to achieve when there is a larger 
proportion of vertices with an even number of neighbours.
The simulation indicates that local properties need not be identical
to those of pure Euclidean gravity, even though we are still
dealing with a summation over (in this case a restricted set of) 
triangulations.

This observation is backed up by considering the dual viewpoint
of equ.(\ref{action}), where one regards the $S^3$ and $S D^2$
terms in equ.(\ref{action}) as representing cubic vertices rather than triangles.
This recasts the model as an $N=1$
version of the $O(N)$ loop gas on a random surface,
whose critical behaviour has been investigated in some detail \cite{6}.
Viewed in this way the model generates $D$ loops floating in
a background of $S$ edges. In this
model, as well as the 
pure Euclidean gravity phase observed
on increasing $g$ with $z$ fixed,
a dense critical phase is encountered
on increasing $z$ with $g$ fixed. In 
this the entire lattice is densely filled with $D$
loops, which translates back to 
the Lorenztian sector of perfectly matched triangulations in
our original, non-dual, picture
\footnote{The coalescence of the two singularities
is the standard Ising model critical point.}.

The $N$ appearing in the $O(N)$ model can conveniently
be parametrized as
\begin{equation}
N = 2 \cos ( \theta \pi ) , \; \; 0 \le \theta \le 1
\end{equation}
Looking at the scaling of various operators and demanding
consistency with the continuum KPZ/DDK \cite{7} results
then gives (for spherical topology on the underlying
Euclidean lattice) that
\begin{eqnarray}
\gamma_{string} &=& - \theta \; \; \; \;   (Critical \; Point) \nonumber \\
\gamma_{string} &=& - { \theta \over 1 - \theta} \; \; \; \;  (Dense \; Phase)
\end{eqnarray}
so in our case $N=1$ gives $\theta = 1/3$. We recover the Ising
model value of $\gamma_{string}= - 1 / 3$ at the critical point
and we find the (Euclidean) pure gravity
value of $\gamma_{string}= - 1 / 2$ throughout the dense
phase that we have identified as Lorentzian.

Although $\gamma_{string}= - 1 / 2$ in the dense/Lorenztian phase
at least some of the local geometrical properties
are not the same as the Euclidean theory.
Watermelon operators
on the form
\begin{equation}
\chi_L = \left< \left( tr ( D )^L \right)^2 \right>,
\end{equation}
measure the expectation values of $L$ $D$ lines
pinned at two points. Their conformal dimensions $\Delta_L$ are known to satisfy
\begin{eqnarray}
\Delta_L &=& {L \over 4} ( 1 + \theta) - {\theta \over 2}  \; \; \; \;
(Critical \; Point) \nonumber \\
\Delta_L &=& {L \over 4} - {1 \over 2} { \theta \over 1 - \theta} \; \; \; \;
(Dense \; Phase)
\end{eqnarray}
As pure gravity is retrieved at the
critical point for $\theta= 1/2$
and Ising link gravity still has $\theta=1/3$ 
in the dense phase these scaling
dimensions are not the same.

\section{Conclusions}

We have suggested that a model 
combining dynamical triangulations
and Ising link quantum gravity 
might capture
some of the features of the discretized microgeometry of a 2D gravity theory 
admitting both Lorentzian and Euclidean signatures.
With suitable restrictions on the allowed triangles
this model is precisely the Ising model coupled to 2D gravity.
The purely Lorentzian sector appears as an antiferromagnetic limit in
the Ising model, giving either a summation over perfectly
matched triangulations or a mixed triangulation/quadrangulation
depending on the choice of convention for the Lorentzian triangles.
As the model is being used away from the Ising critical point some 
properties are essentially those of discretized pure Euclidean gravity
based on triangulations (ie $\gamma_{string}= - 1 /2$), but non-universal features
such as curvature distributions may be different, as witnessed by the simulation and 
the watermelon dimensions.

We have not really resolved the issue of ``factors of $i$''
in this paper. As noted, this
does not play a role in microcanonical simulations but would have
to be faced in extracting scaling information from the results.
However, it is reassuring that the models of 
equ.(\ref{actionIII}) and equ.(\ref{actionV})
representing the extreme Lorentzian limit for both possible
choices of convention {\it can} be
defined for complex couplings, so the standard scaling limit 
can be approached through complex coupling values, which
is precisely what is required in our context.
Modulo this caveat, the results here appear to be rather reassuring
for Euclidean practitioners, as we are still dealing with a theory
of triangulations, albeit decorated, in our toy Lorentzian model.

Perhaps the most important question we have left unanswered in the current
work is whether the microscopic causality structure in our model,
which looks very chaotic, becomes tamer at larger scales. This would
be a prerequisite for extracting a sensible continuum theory.
A numerical investigation of the role of signature changing
would also be possible within the context of the model.
Other ways of grafting a Lorentzian stucture onto Euclidean
lattices also suggest themselves, such as using $\phi^4$ graphs
and taking the edges emanating from a vertex as the lightcone
\cite{seagull}. These might also be amenable to numerical investigation.
In any event the current work suggests that the Ising antiferromagnet,
or more generally the complex temperature Ising model, coupled to
2D gravity might be worthy of further investigation.
 
\section{Acknowledgements}

WB is supported in part by Fonds zur F\"orderung der wissenschaftlichen
Forschung under Contract P9522-PHY.
DJ is partially supported by
EC HCM grant CRRX-CT930343.
WB and DJ would also like 
to acknowledge a British Council/ARC 
exploratory travel grant and the (excellent!) 
hospitality of Harald Markum at the Technical
University of Vienna that made possible the work
in this paper.

\clearpage \newpage
\begin{figure}[htb]
\vskip 20.0truecm
\includegraphics{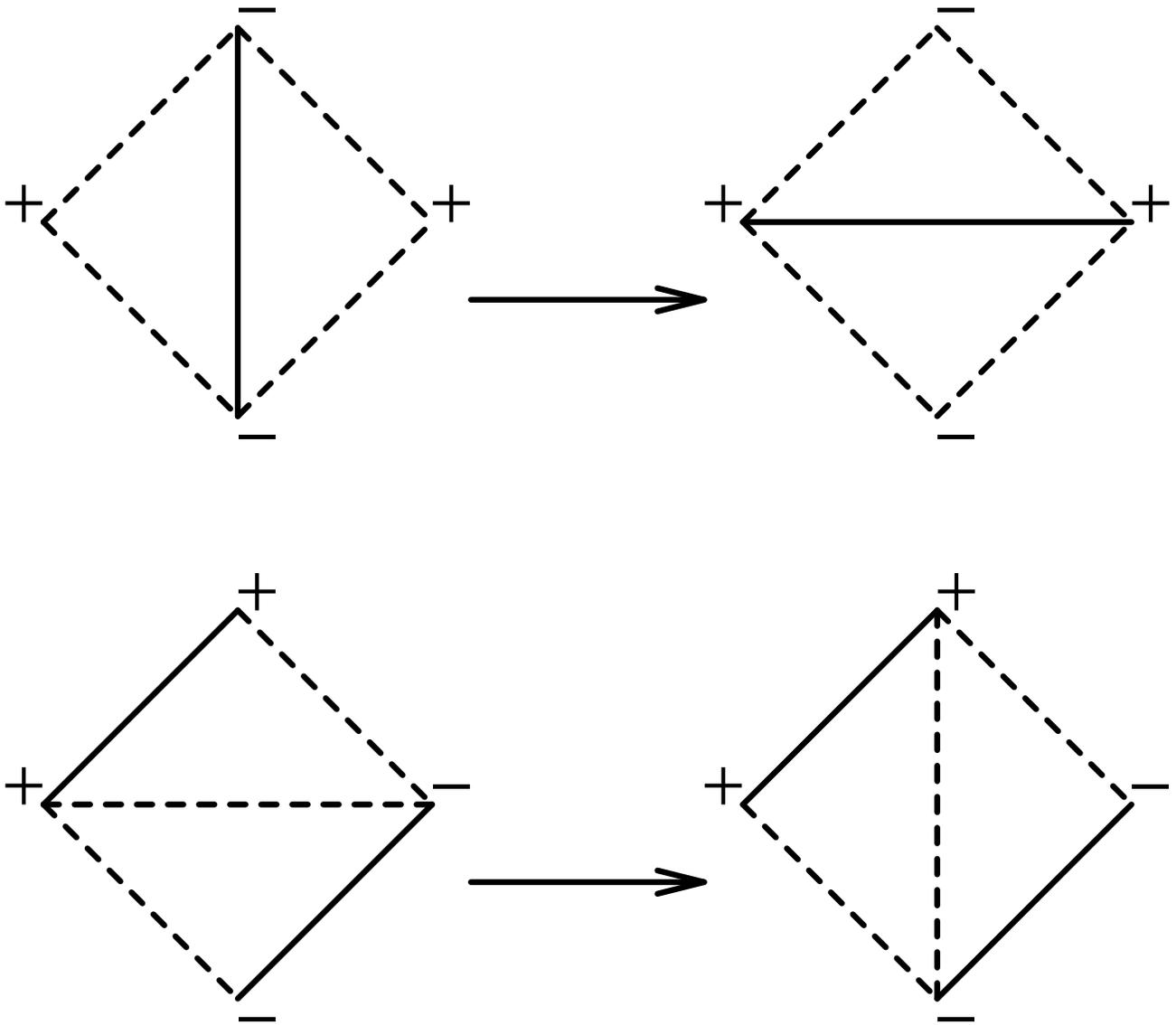}
\caption[]{\label{fig1a}The allowed flip moves
in the purely Lorentzian sector. The timelike edges
are shown dotted and spacelike edges solid. The spins
in the underlying Ising model are also shown.}
\end{figure}
\clearpage \newpage
\begin{figure}[htb]
\vskip 20.0truecm
\includegraphics{fig2a.ps}
\caption[]{\label{fig2a}The two signature changing flip moves.}
\end{figure}
\clearpage \newpage
\begin{figure}[htb]
\vskip 20.0truecm
\includegraphics{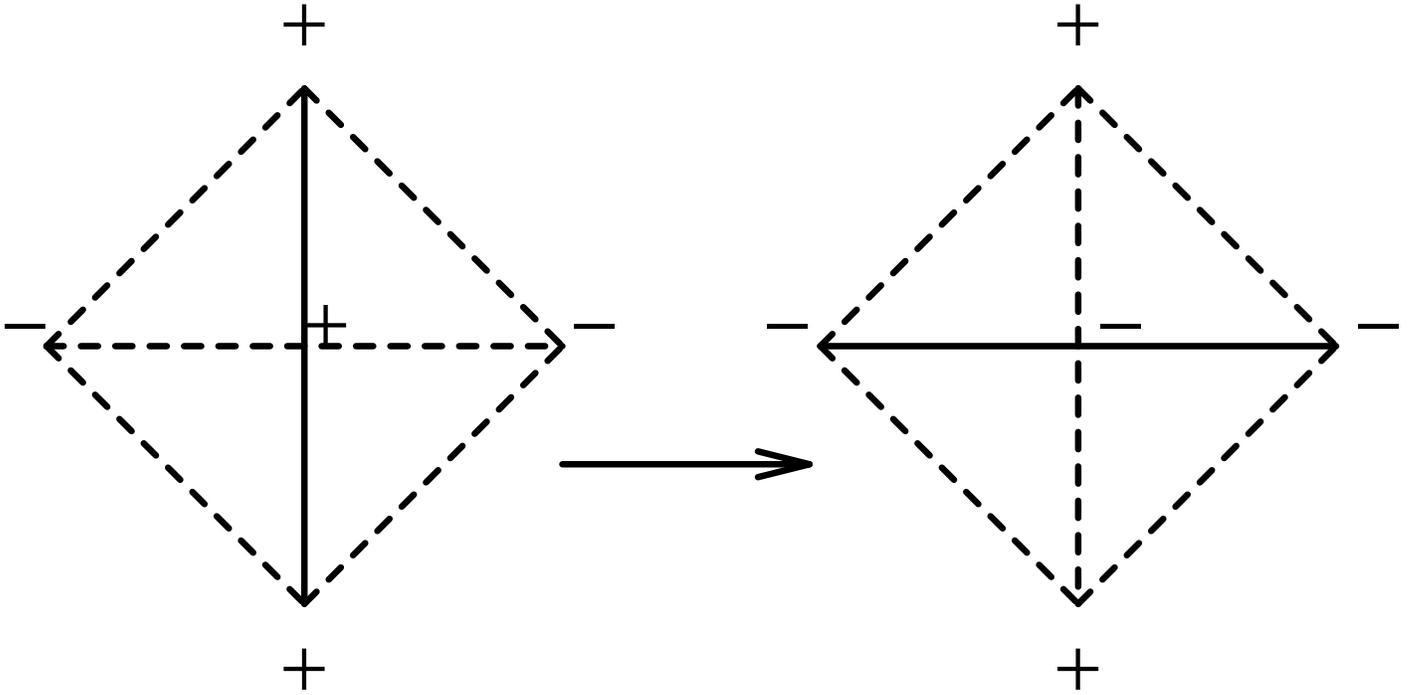}
\caption[]{\label{fig3a}A geometry changing spin
flip in the purely Lorentzian sector.}
\end{figure}
\clearpage \newpage
\begin{figure}[htb]
\vskip 20.0truecm
\includegraphics{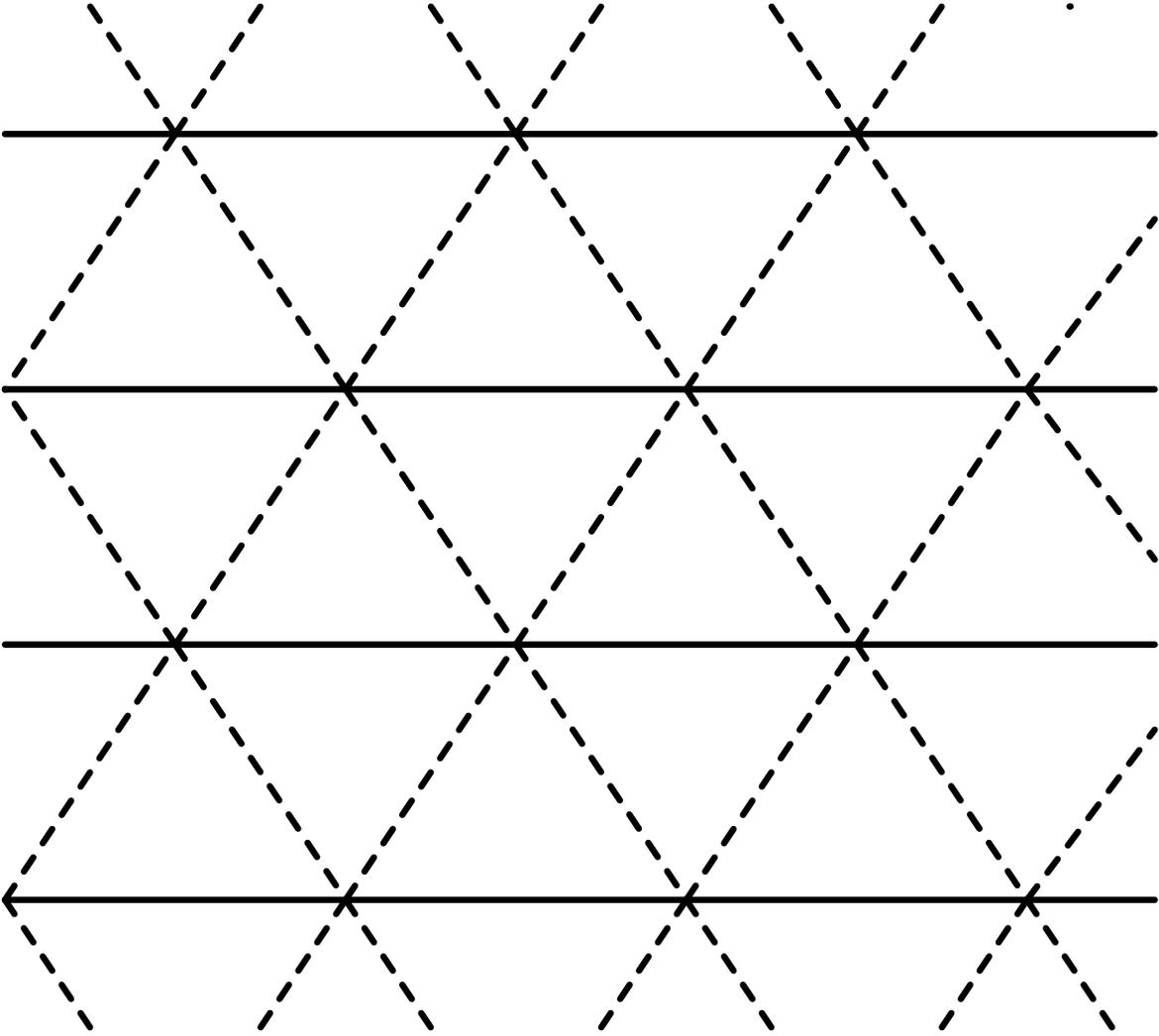}
\caption[]{\label{fig4a}A possible starting configuration in
the purely Lorentzian sector.}
\end{figure}


\bigskip
\begin{thebibliography}{99}
\bibitem{0} For a review see ``Quantization of Geometry'' by J. Ambj\o rn,
Les Houches Summer School Session LXII.
\bibitem{1} T. Regge, Nuovo Cimento 19 (1961) 558.
\bibitem{2} M. Gross and H. Hamber, Nucl. Phys. {\bf B364} (1991) 703;\\
            C. Holm and W. Janke, Phys. Lett. {\bf B335} (1994) 143;\\
             Nucl. Phys. {\bf B} (Proc Suppl.) 42 (1995) 722,725;\\
            ``Measuring the String Susceptibility in 2D Simplicial Quantum
              Gravity using the Regge approach'', hep-lat/9511029.
\bibitem{BB} B. Berg and W. Beirl, Nucl. Phys. {\bf B452} (1995) 415. 
\bibitem{2a} J. Ambj\o rn and J. Jurkiewicz, Nucl. Phys. {\bf B451} (1995) 
643;\\
             J. Ambj\o rn and J. Jurkiewicz, Phys. Lett. {\bf B335} (1995) 
355;\\
             S. Catterall, J. Kogut and R. Renken, Phys. Lett. {\bf B328} 
(1994) 277;\\
             B. de Bakker and J. Smit, Nucl. Phys. {\bf B454} (1995) 343.
\bibitem{2b} W. Beirl, H. Markum and J. Riedler, Phys. Rev. D49 (1994) 5231;\\ 
             B. Berg, Phys. Rev. Lett. {\bf 55} (1985) 904;\\
             Phys. Lett. {\bf B176} (1986) 39;\\
             H. Hamber, Nucl. Phys. {\bf B400} (1993) 347.
\bibitem{3} W. Beirl, H. Markum and J. Riedler, Int. J. Mod. Phys. {\bf C5} 
(1994) 359;\\
            W. Beirl, P. Homolka, B. Krishnan, H. Markum and J. Riedler,
  Nucl. Phys. {\bf B} (Proc Suppl.) 42 (1995) 710;\\
            T. Fleming, M. Gross and R. Renken, Phys. Rev. {\bf D50}
  (1994) 7363.
\bibitem{4a} S. Carroll, M. Ortiz and W. Taylor IV,
            ``A geometric approach to free variable loop equations
              in discretized theories of 2D gravity'', hep-th/9510199;\\
            S. Carroll, M. Ortiz and W. Taylor IV,
            ``Spin/disorder correlations and duality in the $c=1/2$ string'', 
hep-th/9510208.
\bibitem{4} D. Johnston, Phys. Lett. {\bf B314} (1993) 69.
\bibitem{5} V. A. Kazakov, Phys. Lett. {\bf A119} (1986) 140.\\
             D.V. Boulatov and V.A. Kazakov, Phys. Lett. {\bf B186} (1987) 
379.
\bibitem{bachas} C. Bachas, ``On triangles and squares'', CERN-TH.5290/90.
\bibitem{staud} M. Staudacher, Nucl. Phys. {\bf B336} (1990) 349. 
\bibitem{david} F. David, Nucl. Phys. {\bf B348} (1991) 507.
\bibitem{RP} T. Ponzano and T. Regge, ``Semiclassical limit of Racah
             coefficients'', in Spectroscopic and group theoretical methods
             in physics, ed. F. Bloch (North Holland, Amsterdam, 1968).
\bibitem{BF} J. Barrett and T. Foxon, Class. Quant. {\bf 11} (1994) 543.
\bibitem{6} B. Duplantier and I. Kostov, Phys. Rev. Lett. {\bf 61}
            (1988) 1433;\\
            I. Kostov, Mod. Phys. Lett. {\bf A4} (1989) 217;\\
            M. Gaudin and I. Kostov, Phys. Lett. {\bf B220} (1989) 200;\\
            B. Duplantier and I. Kostov, Nucl. Phys. {\bf B340} (1990)
            491;\\
            I. Kostov and M. Staudacher,
            Nucl. Phys. {\bf B384} (1992) 459;\\
            B. Eynard and J. Zinn-Justin, Nucl. Phys. {\bf B386} (1992)
            558;\\
            B. Eynard and C. Kristjansen, Nucl. Phys. {\bf B455} (1995)
            577;\\
            B. Eynard and C. Kristjansen, ``More on the exact solution
            of the $O(n)$ model on a random lattice and an investigation
            of the case $|n|>2$'', SPhT-95/133.
\bibitem{7} V. Khizhnik, A. Polyakov and A. Zamalodchikov, Mod. Phys. Lett.
            {\bf A3} (1988) 819;\\
            J. Distler and H. Kawai, Nucl. Phys. {\bf B321} (1989) 501;\\
            F. David, Mod. Phys. Lett.
            {\bf A3} (1988) 1651.
\bibitem{seagull} W. Siegel, ``Actions for QCD-Like Strings'', hep-th/9601002.
\end{thebibliography}
\end{document}